\documentclass[showpacs,preprintnumbers,amsmath,amssymb,superscriptaddress,11pt]{revtex4}

\usepackage{graphicx}
\usepackage{dcolumn}
\usepackage{bm}

\def\ra{\rangle}
\def\la{\langle}

\def\sin{{\rm sin}}
\def\Tr{{\rm Tr\,}}

\begin{document}
\quad 
\vspace{-2cm}

\begin{flushright}
UT-07-23
\end{flushright}

\vspace*{0.5cm}

\title{Entanglement Entropy in the Calogero-Sutherland Model}

\author{Hosho Katsura}
\email{katsura@appi.t.u-tokyo.ac.jp}
\affiliation{Department of Applied Physics, University of Tokyo,
7-3-1, Hongo, Bunkyo-ku, Tokyo 113-8656, Japan}

\author{Yasuyuki Hatsuda }
\email{hatsuda@hep-th.phys.s.u-tokyo.ac.jp}
\affiliation{Department of Physics, Faculty of Science, University of Tokyo,
7-3-1, Hongo, Bunkyo-ku, Tokyo 113-0033, Japan}

\date{\today}

\begin{abstract}
We investigate the entanglement entropy between two subsets of particles in the ground state of the Calogero-Sutherland model. By using the duality relations of the Jack symmetric polynomials, we obtain exact expressions for both the reduced density matrix and the entanglement entropy in the limit of an infinite number of particles traced out. 
From these results, we obtain an upper bound value of the entanglement entropy. This upper bound has a clear interpretation in terms of fractional exclusion statistics.
\end{abstract}

\maketitle
\section{Introduction}
Entanglement properties of quantum many-body systems have been attracting much attention in quantum information theory and condensed matter physics. 
The entanglement entropy (EE), i.e., the von Neumann entropy of the reduced density matrix of a subsystem, is a measure to quantify how much entangled a many-body ground state is. 
Recently the EE has been used to investigate the nature of quantum ground states such as the quantum phase transition and topological order \cite{Vidal, Levin, Kitaev, Ryu, YH1}. When we  study the entanglement properties in many-body systems, exactly solvable models in one dimension such as the harmonic chain \cite{Eisert}, the XY spin chain in a transverse magnetic field \cite{Vidal, Peschel, Its} and the Affleck-Kennedy-Lieb-Tasaki model \cite{AKLT, Fan, Katsura, Hirano} serve as a laboratory to test the validity of this new concept. The relation between the EE in solvable models and the conformal or massive integrable field theories is extensively discussed in Refs. \cite{Calabrese, Korepin}.

In this article we study the EE of the ground state of the Calogero-Sutherland (CS) model \cite{Calogero, Sutherland}. The CS model is a quantum integrable model with inverse-square interactions on a circle. 
An infinite number of conserved quantities which characterize the integrable structure of this model have been constructed in a Lax form \cite{Ujino}.
Although it is usually a formidable task to compute the correlation functions even in the integrable models \cite{BIK}, one can derive exact expressions for the dynamical correlation functions in this model \cite{Ha, Minahan, Lesage}. This is an important feature of this model which distinguishes itself from the other integrable models. Another interesting aspect of this model is a connection with the fractional statistics in low dimensions. In fractional quantum Hall systems, the ground state wave function is given by the Laughlin state \cite{Laughlin}, and its excitations have fractional charges. Similarly, the ground state of the CS model is described by the Jastrow-type wave function and its excitations are also {\it quasiholes} with fractional charges. Then we can identify the CS model as a canonical model to study the exotic properties of the fractional statistics in low dimensions. It should  be noted here that the EE of the Laughlin state itself is also extensively studied recently \cite{Schoutens, Latorre, Rezayi}.

We consider the EE between two subsystems in the ground state of the CS model. 
Let us first explain how to partition our total system into two subsystems. There are mainly two possible ways to partition the system under consideration. 
One way is to divide the system into two spatial blocks the other to divide the $N$-particle system into an $L$-particle block and an $(N-L)$-particle block. 
They are called a {\it spatial partitioning} and a {\it particle partitioning}, respectively. 
In this article, we focus on the latter. As the EE between two spatial regions in the fractional quantum Hall states can extract a topological quantity such as the total quantum dimension \cite{Schoutens}, the EE based on the particle partitioning in the CS model reveals a new aspect of low-dimensional systems with fractional exclusion statistics. 
First we consider the $L$-particle reduced density matrix of our system. 
By using duality relations of the Jack polynomials, we can formally obtain the exact expression for the reduced density matrix. Although we have the exact form of the reduced density matrix, it is difficult to evaluate the eigenvalues since there are many off-diagonal elements. Then we consider the thermodynamic limit and find that a great simplification occurs in this limit. We should note here that what we mean by the {\it thermodynamic limit} is $(N-L)\to \infty$ limit, where $(N-L)$ is the number of particles traced out. It is slightly different from the usual sense such as $N\to\infty$ with fixed $L/N$. Finally, we focus on the upper bound value of the EE. 
In the thermodynamic limit, we can approximate the reduced density matrix by a maximally entangled state and hence we can evaluate the upper bound by counting the allowed Young tableaux in the duality relation.
The upper bound value is estimated as ${\cal S}^{\rm bound}_{N.L}=\log\binom{\beta(N-L)+L}{L}$ and has a clear interpretation in terms of exclusion statistics \cite{Haldane}. 
We also find that the subleading term of the EE is independent of the total number of particles $N$. 

The organization of this article is as follows.
In Section~\ref{sec:CS}, we will introduce some basic concepts in the CS model used in later sections.
Section~\ref{sec:EE} is the main part of this article. We will calculate the reduced density matrix in the CS model and show that it becomes very simple if we take a thermodynamic limit. Then we will be able to obtain the EE in this limit and to estimate the upper bound of this EE. We will discuss the physical interpretation of this upper bound.
Section~\ref{sec:summary} will be devoted to summary and discussions. 
In Appendix~\ref{app:detail}, we will analyze the EE in the thermodynamic limit more in detail than Section~\ref{sec:EE}.

\section{Calogero-Sutherland model and Jack symmetric polynomials}\label{sec:CS}
\subsection{Calogero-Sutherland model}
\begin{figure}
\includegraphics[width=8cm,clip]{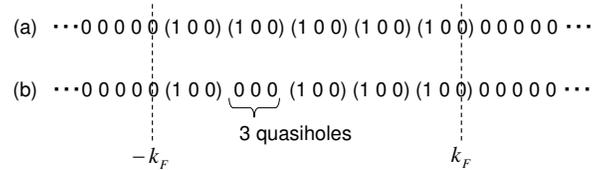}
\caption{(a) The ground state configuration of the quasi-momenta at $\beta=3$ with 5 particles. (b) The excited state with 4 particles and 3 quasiholes obtained by removing one pariticle from the Fermi Sea.}
\label{FermiSea}
\end{figure}

We introduce a precise definition of the CS model. 
The CS model describes the interaction of $N$ particles on a circle of length $l$ and the Hamiltonian is given by
\begin{equation}
H_{CS} = -\sum_{j=1}^N \frac{1}{2} \frac{\partial^2}{\partial x_j^2} + \sum_{i<j} \frac{\beta (\beta-1){(\frac{\pi}{l}})^2}{\sin ^2(\frac{\pi}{l} (x_i-x_j))},
\label{Ham}
\end{equation}
where $x_j$ ($0 \le x_j \le l$) are the coordinates. Here it is convenient to introduce new coordinates
on a unit circle $z_j={\rm exp}(\frac{2 \pi i}{l} x_j)$.  Using these new variables, the exact ground state of $H_{CS}$ is given by the Jastrow-type wave function as 
\begin{equation}
\psi_0 (z_1.z_2, ..., z_N)=\frac{1}{\sqrt{N !}} (\prod_{j=1}^N z_j)^{-\beta \frac{N-1}{2}}
\prod_{i<j} (z_i-z_j)^\beta.  	
\label{gswf}
\end{equation}
All the excited states of this model can also be obtained by multiplying certain symmetric polynomials to $\psi_0$ as
\begin{equation}
\psi_{\lambda}(z_1,z_2, ..., z_N)=P_{\lambda}(z_1,z_2, ..., z_N;\beta) \psi _0(z_1.z_2, ..., z_N).
\label{excite}
\end{equation}
The symmetric polynomial in Eq. ({\ref{excite}}) are called the Jack symmetric polynomials and
characterized by partitions $\lambda$.
The partition $\lambda$ is a sequence $\lambda=(\lambda_1,\lambda_2,..., \lambda_r, ...)$ of non-negative integers in decreasing order: $\lambda_1 \ge \lambda_2 \ge ... \ge \lambda_r \ge ...$\hspace{1mm}. 
Let us introduce some terminology. We use the notation of Macdonald \cite{Macdonald}. Every partition has a corresponding {\it Young tableau} which graphically represents a partition (see Fig.~\ref{young}).
The non-zero $\lambda_i$ are called the {\it parts} of $\lambda$. The number of parts is the {\it length} of $\lambda$, denoted by $l(\lambda)$ and the sum of the parts is the {\it weight} of $\lambda$ denoted by $|\lambda|$ and explicitly written as $|\lambda|=\sum^{l(\lambda)}_{i=1}\lambda_i$. 
The excitation energy is also characterized by the partition as
\begin{equation}
E_\lambda = \frac{1}{2}\Big(\frac{2\pi}{l}\Big)^2 \sum^N_{i=1}k^2_i(\lambda),
\end{equation} 
where the quasi-momentum $k_i(\lambda) = \lambda_i+\beta(\frac{N+1}{2}-i)$.
The set of quasi-momenta is subject to the exclusion constraint $k_i-k_{i+1}\ge \beta$. In the ground state, the configuration of the quasi-momenta is given by $k_i(0)=\beta(\frac{N+1}{2}-i)$ and this configuration is schematically shown in Fig.~\ref{FermiSea}.(a). We call this configuration the Fermi sea.  
In Fig.~\ref{FermiSea}, a {\it particle} can be identified by one 1 followed by $\beta-1$ zeros and a {\it quasihole} by one 0. Therefore, if we remove $n$ particles from the Fermi sea, $\beta n$ quasiholes are created in the Fermi sea (see Fig.~\ref{FermiSea}.(b)). We should note here that the coupling $\beta$ has been assumed to be a positive integer for the sake of simplicity in this article. However, in principle, we can extend this correspondence at any positive rational coupling $\beta=p/q$ \cite{Serban}.
\subsection{Jack symmetric polynomials}
\begin{figure}
\includegraphics[width=7cm,clip]{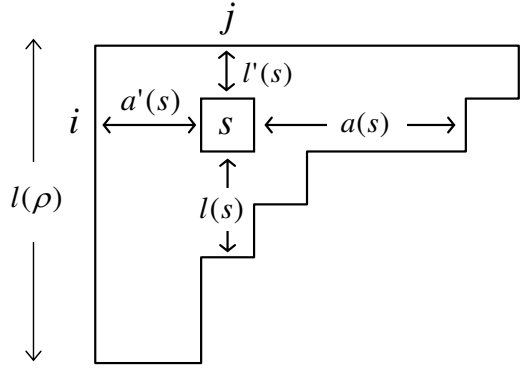}
\caption{The young tableau corresponding to the partition $\lambda$. The box in the tableau indicates $s=(i,j)$. $a(s), l(s),a'(s)$ and $l'(s)$ are arm-length, leg-length, arm-colength and leg-colength, respectively.  }
\label{young}
\end{figure}

Let us turn to focus on the mathematical aspects of the Jack symmetric polynomials. The Jack symmetric polynomials are mutually orthogonal with respect to the following scalar product on the ring of symmetric polynomials in $N$ indeterminates $z_1, ..., z_N$:
\begin{equation}
\langle f,g \rangle^\prime _N = \oint \frac{dz_1}{2\pi i z_1} \cdots \oint \frac{dz_N}{2 \pi i z_N}
\overline{f(z_1, z_2, ... ,z_N)}  g(z_1, z_2,..., z_N) |\psi_0 (z_1, z_2, ..., z_N)|^2.
\label{inner}
\end{equation} 
The normalization of the ground state wave function $\psi_0$ is defined as
${\cal N}(\beta, N)=\la 1,1\ra_N'$ and its explicit form is given by
$\frac{(N \beta)!}{(\beta !)^N N!}$.
The explicit orthogonality relation for the Jack polynomials is given by
\begin{equation}
\langle P_{\lambda}, P_{\mu} \rangle^\prime _N =\delta_{\lambda,\mu} {\cal N}(\beta, N)
\prod_{s \in \lambda}\frac{a(s)+\beta l(s) +1}{a(s)+\beta l(s) +\beta} \prod_{s \in \lambda} \frac{\beta N +a'(s)-\beta l'(s)}{\beta N+a'(s)+1-\beta(l'(s)+1)},
\end{equation}
where $s=(i,j)$ is a box on a Young tableau identified by its coordinates $1 \le i \le l(\lambda)$ and $1 \le j \le \lambda_i$. 
The notations $a(s), l(s), a'(s)$ and $l'(s)$ are summarized in Fig. \ref{young}. 
It is well known that classical families of symmetric polynomials can be obtained by specializing the coupling $\beta$ of the Jack symmetric polynomials. 
For $\beta=0,\,1,\,2$, and $\infty$, the Jack symmetric polynomials are  reduced to the monomial symmetric, the Schur, the zonal, and the elementary symmetric polynomials, respectively \cite{Macdonald}. 

\section{Reduced density matrix and entanglement entropy} \label{sec:EE}
\begin{figure}
\includegraphics[width=7cm,clip]{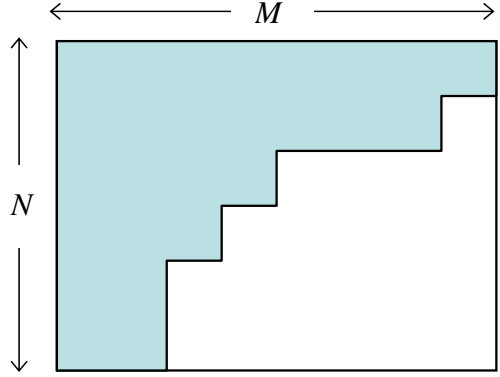}
\caption{Young tableaux within the shaded region are allowed in the expansion formula (\ref{duality}).}
\label{allowed}
\end{figure}

In this section, we consider the reduced density matrix and the entanglement entropy for any subset of $L$ particles in a system of $N$ particles in the state (\ref{gswf}). 
The $L$-particle reduced density matrix, being normalized, is defined as
\begin{eqnarray}
&&\rho(\overline{w}_1, ..., \overline{w}_L; z_1, ..., z_L) = \nonumber \\
&&\frac{1}{{\cal N}(\beta, N)} \oint \frac{dz_{L+1}}{2 \pi iz_{L+1}}\cdots \oint \frac{dz_N}{2 \pi i z_N} \overline{\psi_0 (w_1,...,w_L, z_{L+1}, ..., z_N)} \psi_0 (z_1,...,z_L, z_{L+1}, ..., z_N ).
\label{RDM1}
\end{eqnarray}
Here the partial trace is taken over the variables $z_{L+1}, ...,z_N$. 
To calculate the EE, it is useful to introduce a trace in a complex integral form. The trace of any $L$-particle operator $A(\overline{w}_1, ..., \overline{w}_L; z_1, ..., z_L)$ is defined by
\begin{equation}
\Tr [A] \equiv
\oint \frac{dz_1}{2 \pi iz_1}\cdots \oint \frac{dz_L}{2 \pi i z_L}
A(\overline{z}_1, ..., \overline{z}_L;z_1, ..., z_L).
\end{equation}
Since the reduced density matrix (\ref{RDM1}) is normalized, $\Tr [\rho]=1$.
Similarly, the trace of the product of any $L$-particle operators $A(\overline{w}_1, ..., \overline{w}_L; z_1, ..., z_L)$ and $B(\overline{w}_1, ..., \overline{w}_L; z_1, ..., z_L)$ is defined by
\begin{equation}
\Tr [AB] \equiv 
\oint \frac{dw_1}{2 \pi iw_1}\cdots \oint \frac{dw_L}{2 \pi i w_L}
\oint \frac{dz_1}{2 \pi iz_1}\cdots \oint \frac{dz_L}{2 \pi i z_L}
A(\overline{w}_1, ..., \overline{w}_L;z_1, ..., z_L) B(\overline{z}_1, ..., \overline{z}_L;w_1, ..., w_L), \nonumber
\end{equation}
and the EE is defined by ${\cal S}_{N,L}=-\Tr[\rho \log \rho]$.
To obtain the explicit form of the reduced density matrix, it is convenient to rewrite Eq. (\ref{RDM1})
by using the ground state wave functions of the subsystems, $\psi_0(z_1, z_2, ..., z_L)$ and $\psi_0(z_{L+1}, ..., z_N)$, as
\begin{eqnarray}
&&\rho(\overline{w}_1, ..., \overline{w}_L;z_1, ..., z_L)=\frac{1}{{\cal N}(\beta,N)} \frac{L! (N-L)!}{N!} \Big( \prod_{i=1}^L \overline{w}_i z_i \Big)^{-\beta\frac{N-L}{2}} \overline{\psi_0 (w_1, ..., w_L)} \psi_0 (z_1, ..., z_L) \nonumber \\
&&\times \oint \frac{dz_{L+1}}{2 \pi i z_{L+1}} \cdots  \oint \frac{dz_N}{2\pi iz_{N}} 
\prod_{i=1}^L \prod_{j=L+1}^N (1-z_i \overline{z}_j)^\beta (1-\overline{w}_i z_j)^\beta |\psi_0 (z_{L+1}, ..., z_N)|^2.
\label{RDM2}
\end{eqnarray} 
Recalling the definition of the scalar product (\ref{inner}), Eq. (\ref{RDM2}) can be rewritten again as
\begin{equation}
\frac{1}{{\cal N}(\beta,N)} \frac{1}{\binom NL} \overline{\Psi_0 (w_1, ..., w_L)} \Psi_0 (z_1, ..., z_L)
\Bigg\langle \prod_{i=1}^L \prod_{j=L+1}^N (1-\overline{z}_i z_j)^\beta,  
           \prod_{i=1}^L \prod_{j=L+1}^N (1-\overline{w}_i z_j)^\beta \Bigg\rangle_{N-L}^\prime,
\label{RDM3}
\end{equation}
where $\Psi_0(z_1, ..., z_L) \equiv (\prod^L _{i=1} z_i)^{-\beta(N-L)/2} \psi_0(z_1, ..., z_L)$.
The next thing to do is to compute the scalar product in Eq. (\ref{RDM3}). 
Let us now introduce the following duality relation to carry out our calculation \cite{Iso, Lesage}:
\begin{equation}
\prod_{i=1}^N \prod_{j=1}^M (1+x_i y_j) = \sum_{\lambda} P_{\lambda}(x_1,x_2,...,x_N;\beta) P_{\lambda'}(y_1,y_2,...,y_M;1/\beta).
\label{duality}
\end{equation}
Here, the conjugate partition $\lambda'$ is a transpose of the Young tableau $\lambda$ and partitions $\lambda$ are summed over the Young tableaux which satisfy $l(\lambda) \le N$ and $l(\lambda') \le M$ (see Fig. (\ref{allowed})). 
The duality relation Eq.(\ref{duality}) plays a crucial role to simplify the reduced density matrix (\ref{RDM3}). We shall explain the procedure of the calculation in more details. First, we introduce dummy variables $z^{(k)}_j$, $(L+1 \le j \le N$, $1 \le k \le \beta)$. Secondly, we expand $\prod^L_{i=1}\prod_{(j,k)}(1-{\overline z}_i z^{(k)}_j)$ by using the duality relation (\ref{duality}). Here, $(j, k)$ runs from $(L+1, 1)$ to $(N, \beta)$. 
Finally, we set the dummy variables $z^{(k)}_j=z_j$, ($1 \le k \le \beta$). 
We can summarize the above as the following expansion formula: 
\begin{equation}
\prod_{i=1}^L \prod_{j=L+1}^N (1-\overline{z}_i z_j)^\beta =
\sum_{\lambda}  P_{\lambda}({\overline z}_1, ..., {\overline z}_L;\beta) P_{\lambda'}(\overbrace{-z_{L+1}, ..., -z_{L+1}}^{\beta}, ..., \overbrace{-z_N, ..., -z_N }^{\beta};1/\beta),
\label{duality11}
\end{equation}
where partitions $\lambda$ are summed over those that satisfy $l(\lambda) \le L$ and $l(\lambda') \le \beta (N-L)$. Here we have also assumed that the coupling $\beta$ is a positive integer.
The above formula has a clear physical interpretation as a superposition of the intermediate states consist of $L$ particles and $\beta(N-L)$ quasiholes. 

Next, we try to rewrite $P_{\lambda^\prime}$ with coupling $1/\beta$ in Eq. (\ref{duality11}) in terms of $P_{\lambda}$ with $\beta$. 
It is also well known that the Jack symmetric polynomials can be expressed as polynomials in
power sums $p_n=\sum_i z^n_{i}$. We give as examples the expressions up to $|\lambda|=3$:
\begin{equation}
\begin{array}{c}
P_{(1)}=p_1, \\
\left\{\begin{array}{l}
P_{(2)}=\frac{1}{1+\beta }p_2 +\frac{\beta}{1+\beta}p^2_1, \\
P_{(1,1)}=-\frac{1}{2}p_2 + \frac{1}{2}p^2_1, \\
\end{array}\right.
\end{array}
\hspace{5mm}
\left\{\begin{array}{l}
P_{(3)}= \frac{2}{(1+\beta)(2+\beta)}p_3 + \frac{3\beta}{(1+\beta)(2+\beta)}p_2 p_1 +\frac{\beta^2}{(1+\beta)(2+\beta)}p^3_1, \\
P_{(2,1)}=-\frac{1}{1+2\beta}p_3 + \frac{1-\beta}{1+2\beta} p_2 p_1 + \frac{\beta}{1+2\beta}p^3_1, \\
P_{(1,1,1)}= \frac{1}{3}p_3-\frac{1}{2}p_2 p_1 +\frac{1}{6}p^3_1.
\end{array}\right.
\end{equation}
We define the Jack symmetric polynomials whose arguments are power sums as
$P^{(\alpha)}_{\lambda}(\{p_n(z_j)\}) \equiv P_{\lambda}(z_{L+1}, ..., z_{N};\beta)$, where $\alpha=1/\beta$.
Another important duality relation between the Jack polynomials with couplings $\beta$ and $1/\beta$ is given by
\begin{equation}
\omega_{\alpha}(P^{(\alpha)}_{\lambda}(\{ p_n \})) = \frac{c^{\prime}_{\lambda}(\alpha)}{c_{\lambda}(\alpha)} P^{(1/\alpha)}_{\lambda'}(\{p_n \}),
\label{duality2}
\end{equation}
where
$c_{\lambda}(\alpha)=\prod_{s \in \lambda} (\alpha a(s)+l(s)+1)$ and
$c^\prime _{\lambda}(\alpha)=\prod_{s \in \lambda} (\alpha a(s)+l(s)+\alpha)$.
In Eq. (\ref{duality2}), $\omega_{\alpha}$ is an {\it involution}, an automorphism on the ring of symmetric polynomials, and is defined by
\begin{equation}
\omega_{\alpha}(p_n)=-(-1)^n \alpha p_n.
\end{equation}
Using the second duality relation Eq. (\ref{duality2}), we can rewrite $P_{\lambda^\prime}$ in Eq. (\ref{duality11}) as
\begin{equation}
P_{\lambda'}(\overbrace{-z_{L+1}, ..., -z_{L+1}}^{\beta}, ..., \overbrace{-z_N, ..., -z_N }^{\beta};1/\beta)=
\frac{c_{\lambda}(\alpha)}{c^\prime _{\lambda}(\alpha)} P^{(\alpha)}_{\lambda}(\{-p_n(z_j) \}).
\label{betabeta}
\end{equation}
We should note here that the argument of $P^{(\alpha)}_{\lambda}$ in the right hand side of Eq. (\ref{betabeta}) is not power-sum $p_n$ itself but $-p_n$ and hence $P_{\lambda}^{(\alpha)}(\{-p_n(z_j) \}) \ne P_{\lambda}(z_{L+1},z_{L+2}, ..., z_N;\beta)$. In other words, $P_{\lambda}^{(\alpha)}(\{-p_n(z_j) \})$ is expanded by the original Jack polynomials $P_{\mu}(z_{L+1},z_{L+2}, ..., z_N;\beta)$ with $|\mu|=|\lambda|$. 
By substituting Eqs.~(\ref{duality11}) and (\ref{betabeta}) into Eq.~(\ref{RDM3}), we formally obtain
\begin{eqnarray}
&&\rho(\overline{w}_1, ..., \overline{w}_L;z_1, ..., z_L)=\frac{1}{{\cal N}(\beta,N)}\frac{1}{\binom NL}
\overline {\Psi_0 (w_1, ..., w_L)} \Psi_0 (z_1, ..., z_L)
\nonumber \\
&&\times \sum_{\lambda_1, \lambda_2} 
\big\la P^{(\alpha)}_{\lambda_1}(\{-p_n(z_j) \}), P^{(\alpha)}_{\lambda_2}(\{-p_n(z_j) \}) \big\ra^\prime_{N-L}
\frac{c_{\lambda_1}(\alpha)}{c'_{\lambda_1}(\alpha)} 
\frac{c_{\lambda_2(\alpha)}}{c'_{\lambda_2}(\alpha)}
P_{\lambda_1}({\overline w}_1, ..., {\overline w}_L;\beta)
P_{\lambda_2}(z_1, ..., z_L;\beta). \nonumber \\
\label{maxent}
\end{eqnarray}
We stress that the form of the reduced density matrix (\ref{maxent}) is exact even when $(N-L)$ is finite. Let us see the structure of the reduced density matrix more closely. Since $\la P^{(\alpha)}_{\lambda_1}(\{-p_n\}),P^{(\alpha)}_{\lambda_2}(\{-p_n \}) \ra'_{N-L}=0$ when $|\lambda_1|\ne|\lambda_2|$, (\ref{maxent}) is block-diagonal and the size of each block is $d(m) \times d(m)$ ($m=0,1,...,\beta(N-L)\times L$), where $d(m)$ is the number of partitions $\lambda$ satisfying $l(\lambda)\le L,\, l(\lambda')\le\beta(N-L)$, and $|\lambda|=m$. Therefore, in principle, we can numerically obtain exact eigenvalues of the density matrix by diagonalizing all the blocks in (\ref{maxent}). Although the original problem is reduced to the finite-dimensional eigenvalue problem, it is also difficult to evaluate the eigenvalues of submatrices when $d(m)$ is large.
However, if we consider $(N-L)\to \infty$ limit, a considerable simplification occurs and we can evaluate the EE without any numerical calculations. 

Let us now consider the thermodynamic limit of the subsystem which traced out, i.e., $(N-L) \to \infty$. The crucial point in our calculation is that $P^{(\alpha)}_{\lambda}(\{-p_n \})$ are asymptotically orthogonal with each other if we take the limit $(N-L) \to \infty$.
In this limit, the reduced density matrix of our subsystem (\ref{maxent}) becomes similar to the {\it maximally entangled state}. To see this, it is useful to expand the Jack symmetric functions in terms of the power sum symmetric functions \cite{Macdonald} as
\begin{equation}
P^{(\alpha)}_{\lambda}(\{ p_n \}) = c_{\lambda}(\alpha)^{-1}\sum_{\rho} \theta^{\lambda}_{\rho}(\alpha) p_{\rho},
\label{expandj}
\end{equation}
where the power sum symmetric functions are defined for a partition $\rho=(\rho_1, \rho_2, ..., \rho_{l(\rho)})$ as $p_{\rho} \equiv \prod^{l(\rho)}_{i=1} p_{\rho_i}$.
The coefficients $\theta^{\lambda}_{\rho}(\alpha)$ satisfy the following orthogonality relations \cite{Macdonald}:
\begin{eqnarray}
\sum_{\rho} z_{\rho} \alpha^{l(\rho)} \theta^{\lambda}_{\rho}(\alpha) \theta^{\mu}_{\rho}(\alpha) = \delta_{\lambda \mu} c_{\lambda}(\alpha) c^\prime_{\lambda}(\alpha) \nonumber \\ 
\sum_{\lambda} c_{\lambda}(\alpha)^{-1} c^\prime _{\lambda} (\alpha)^{-1} \theta^{\lambda}_{\rho} (\alpha) \theta^{\lambda}_{\sigma} (\alpha) = \delta_{\rho \sigma} z_{\rho}^{-1} \alpha^{-l(\rho)},
\label{orthocoeff}
\end{eqnarray}
where $z_\rho=\prod_{i \ge 1}i^{m_i}m_i!$ with $m_i$, the number of parts of $\rho$ equal to $i$.
The coefficients $\theta^{\lambda}_{\rho}(\alpha)$ are nonzero if and only if $|\lambda|=|\rho|$.
From these relations, we can easily expand the power sum symmetric functions $p_{\rho}$ in terms of the Jack symmetric functions as
\begin{equation}
p_{\rho}=\sum_{\mu} z_{\rho} \alpha^{l(\rho)} \theta^{\mu}_{\rho}(\alpha) c^\prime_{\mu}(\alpha)^{-1} P^{(\alpha)}_{\mu}(\{ p_n \}).
\label{expandp}
\end{equation}
We stress here that the above relation itself does not depend on whether we consider the Jack symmetric polynomials in a finite number of variables or the Jack symmetric functions in infinitely many variables.  
By using Eqs. (\ref{expandj}) and (\ref{expandp}), we can formally expand  $P^{(\alpha)}_{\lambda}(\{-p_n\})$ in terms of $P^{(\alpha)}_{\lambda}(\{p_n \})$ as
\begin{equation}
P^{(\alpha)}_{\lambda}(\{-p_n(z_j) \})=c_{\lambda}(\alpha)^{-1} \sum_{\rho} \sum_{\mu} (-\alpha)^{l(\rho)} z_{\rho} \theta^{\lambda}_{\rho}(\alpha) \theta^{\mu}_{\rho}(\alpha) c^{\prime} _{\mu}(\alpha)^{-1} P^{(\alpha)}_{\mu}(\{p_n(z_j) \}).
\end{equation}
Now we are ready to see the asymptotic orthoginality of $P^{(\alpha)}_{\lambda}(\{-p_n \})$.
The scalar product of $P^{(\alpha)}_{\lambda_1}(\{-p_n \})$ and $P^{(\alpha)}_{\lambda_2}(\{-p_n \})$ can be represented as
\begin{eqnarray}
&&\langle P^{(\alpha)}_{\lambda_1}(\{-p_n(\zeta_j)\}), P^{(\alpha)}_{\lambda_2}(\{-p_n(\zeta_j)\}) \rangle^{\prime}_{N-L}
= c_{\lambda_1}(\alpha)^{-1} c_{\lambda_2}(\alpha)^{-1} \nonumber \\
&&\times \sum_{\rho_1,\rho_2 \atop \mu_1, \mu_2} (-\alpha)^{l(\rho_1)+l(\rho_2)} z_{\rho_1} z_{\rho_2} \theta^{\lambda_1}_{\rho_1}(\alpha)
\theta^{\mu_1}_{\rho_1}(\alpha) \theta^{\lambda_2}_{\rho_2}(\alpha)
\theta^{\mu_2}_{\rho_2}(\alpha) c^\prime _{\mu_1}(\alpha)^{-1} c^\prime _{\mu_2} (\alpha)^{-1} 
\langle P^{(\alpha)}_{\mu_1},P^{(\alpha)}_{\mu_2} \rangle^\prime_{N-L}.
\label{-p_n_scalar}
\end{eqnarray}
Suppose that the number of the particles in the subsystem traced out, $(N-L)$, is sufficiently large, i.e., in the thermodynamic limit, we can simplify the scalar product in Eq. (\ref{-p_n_scalar}) as
\begin{equation}
\lim_{N-L \to \infty} \la P^{(\alpha)}_{\mu_1},P^{(\alpha)}_{\mu_2} \ra^\prime_{N-L} =
\delta_{\mu_1 \mu_2} {\cal N}(\beta, N-L) \frac{c^\prime _{\mu_1}(\alpha)}{c_{\mu_1}(\alpha)}.
\end{equation}
In this limit, we can apply the orthogonality relations (\ref{orthocoeff}) to Eq. (\ref{-p_n_scalar}) and hence we obtain
\begin{equation}
\la P^{(\alpha)}_{\lambda_1}(\{-p_n(\zeta_j)\}), P^{(\alpha)}_{\lambda_2}(\{-p_n(\zeta_j)\}) \ra^{\prime}_{N-L}
\sim \delta_{\lambda_1 \lambda_2} {\cal N}(\beta, N-L) \frac{c^\prime _{\lambda_1}(\alpha)}{c_{\lambda_1}(\alpha)}.
\label{-p_n_scalar2}
\end{equation}
We call this relation an {\it asymptotic orthogonality} of $P^{(\alpha)}(\{-p_n \})$. The crucial point in the above calculation is that the sign factor $(-1)^{l(\rho_1)+l(\rho_2)}$ which originally comes from the expansion of $P^{(\alpha)}_{\lambda}(\{-p_n \})$ in Eq. (\ref{-p_n_scalar}) is canceled out. 
By substituting Eq. (\ref{-p_n_scalar2}) into Eq. (\ref{maxent}), the asymptotic form of the reduced density matrix can be expressed by the normalized basis ${\tilde P}_\lambda$ as
\begin{equation}
\rho(\overline{w}_1, ..., \overline{w}_L; z_1, ..., z_L)
\sim \sum_{\lambda} D_{\lambda} 
\tilde P_{\lambda}(\overline{w}_1, ..., \overline{w}_L;\beta) \tilde P_{\lambda} (z_1, ..., z_L;\beta) \overline{\Psi_0(\{w_j \})} \Psi_0(\{z_j \}),
\label{maxent2}
\end{equation}
where $D_\lambda$ and $\tilde P_\lambda$ are defined as
\begin{equation}
D_\lambda=\frac{1}{\binom {N\beta}{L\beta}}
\prod_{s \in \lambda}\frac{\beta L +a'(s)-\beta l'(s)}{\beta L+a'(s)+1-\beta (l'(s)+1)}
\label{coeff_D}
\end{equation}
and $\tilde P_\lambda = P_\lambda/\sqrt{ \la P_\lambda, P_\lambda\ra'_L }$,
respectively.
Then we can obtain an exact expression for the EE in the thermodynamic limit as 
\begin{equation}
{\cal S}_{N,L}=-\sum_\lambda D_\lambda \log D_\lambda.
\label{eq:EE}
\end{equation}
Although this is the exact expression for the EE in the large-$(N-L)$ limit, it is formidable to sum up all $D_\lambda \log D_\lambda$ because they depend on $\lambda$ in a complicated way. To see the physical meaning of this value, let us now evaluate the upper bound value of the EE. Since the reduced density matrix $\rho(\overline{w}_1, ..., \overline{w}_L; z_1, ..., z_L)$ has already been normalized, we immediately notice that $\Tr \rho=\sum_\lambda D_\lambda=1$. Under this constraint, $-\sum_\lambda D_\lambda \log D_\lambda$ takes the maximum value when all $D_\lambda$'s are equal. We can take this maximum value as the upper bound. This maximization corresponds to neglecting the fact that $D_\lambda$ depends on the shape of the Young tableau. From the viewpoint of quantum information, we can say that the reduced density matrix (\ref{maxent2}) can be approximated by a {\it maximally entangled state}. The upper bound value of the EE is completely determined by the number of allowed tableaux.
Since the allowed partitions in the duality expansion Eq. (\ref{duality11}) satisfy $l(\lambda) \le L$ and $l(\lambda') \le \beta (N-L)$, the total number of allowed tableaux is easily obtained as $\binom{\beta (N-L)+L}{L}$. Then the upper bound value of the EE is given by
\begin{equation}
{\cal S}_{N,L} \le {\cal S}_{N,L}^{\rm bound} = \log \binom{\beta (N-L)+L}{L},
\end{equation}
where the equality holds when $\beta=1$, i.e., the free-fermion case. 
Although it is one of the general properties that the EE is invariant under the replacement $L \to N-L$, $N-L \to L$, the upper bound itself does not satisfy this property: ${\cal S}_{N,L}^{\rm bound}\ne {\cal S}_{N,N-L}^{\rm bound}$. 
This fact means that ${\cal S}_{N,N-L}$ approaches ${\cal S}_{N,L}^{\rm bound}$, not ${\cal S}_{N,N-L}^{\rm bound}$ when $(N-L)\to \infty$. 
The upper bound ${\cal S}_{N,L}^{\rm bound}$ enables us to understand the physical meaning of the EE in the ground state of the CS model.
We now try to explain it in terms of exclusion statistics. In the ground state of the CS model, occupied quasi-momenta $k_i(0)$ are separated by $\beta-1$ unoccupied ones. We can schematically describe this configuration as Fig.\ref{FermiSea}(a). In our calculation of the EE, tracing out one particle from the $N$-particle ground state corresponds to the decimation of one quasi-momentum from the Fermi sea. In other words, one 1 is removed from the Fermi sea when we trace out one of coordinates $z_j$. As we said before, $\beta$ quasiholes ($\beta$ zeroes) are created in the Fermi sea in this process (see Fig. \ref{FermiSea}(b)). It is now obvious that tracing out $(N-L)$ particles from the ground state corresponds to the decimation of $(N-L)$ quasi-momenta from the Fermi sea and the creation of $\beta(N-L)$ quasiholes in the Fermi sea. The number of possible intermediate states consisting of $L$ particles and $\beta (N-L)$ quasiholes can be counted as follows. First we recall that the Fermi sea consists of $N$ 1's and $(\beta-1)N$ 0's. After the decimation of the $(N-L)$ quasi-momenta, the configuration of the state consists of $L$ 1's and $(\beta-1)N+(N-L)$ 0's with the exclusion constraint such that any two 1's are separated by more than $(\beta-1)$ 0's. Finally, we notice that the number of possible intermediate states is identical to that of possible configurations of 1's and 0's with the constraint and can be easily obtained as $\binom{\beta(N-L)+L}{L}$. Here we can see that the upper bound of the EE ${\cal S}_{N,L}^{\rm bound}$ is equal to the logarithm of this number. 
It is also remarkable that ${\cal S}_{N,L}^{\rm bound}$ coincides with the upper bound value of the EE in the Laughlin state if we identify $m=\beta$, where $m$ denotes the inverse of the filling factor $\nu$ \cite{Rezayi}. It would also be possible to interpret ${\cal S}_{N,L}^{\rm bound}$ in terms of the {\it flux attachment} in the context of the quantum Hall effect. 
While ${\cal S}_{N,L}^{\rm bound}$ provides a natural way to understand the EE in the CS model in terms of the fractional exclusion statistics, we can also obtain a more accurate value of the EE by taking it into account that $D_\lambda$ depends on the shape of the Young tableau $\lambda$. 
Comparing this value with ${\cal S}_{N,L}^{\rm bound}$, we notice that the subleading term, ${\cal S}_{N,L}-{\cal S}_{N,L}^{\rm bound}$, does not depend on the total number of particles $N$ but only on the coupling $\beta$ and $L$. A similar universal property has already been found in the study of the one-particle EE of hard-core anyons on a ring, where the subleading term depends only on the anyonic parameter $\theta$ \cite{Santachiara}.  
The details of the calculations and the difference between ${\cal S}_{N,L}$ and ${\cal S}_{N,L}^{\rm bound}$ in the thermodynamic limit are argued in Appendix \ref{app:detail}.

\section{Summary and discussions} \label{sec:summary}
In this paper, we have studied the entanglement entropy between two blocks of particles in the ground state of the Calogero-Sutherland model. We have obtained the exact expressions for both the reduced density matrix of the subsystem and entanglement entropy in the limit of an infinite number of particles traced out. In our calculation, the duality relations between the Jack symmetric polynomials with coupling $\beta$ and those with $1/\beta$ have played a crucial role. From the obtained results, we have estimated the upper bound value of the EE by a variational argument. We have also found that the upper bound value itself has a clear physical meaning in terms of fractional exclusion statistics. This interpretation indicates that entanglement between subsets of particles enables us to extract interesting properties in a wide range of systems with fractional exclusion statistics. 
It is also remarkable that this upper bound coincides with that of the Laughlin state in fractional quantum Hall systems when we identify the inverse of the filling factor $m=\beta$.

While we have studied the EE between two blocks of particles, it would also, of course, be important to study the EE between two spatial regions in the ground state of the CS model. In spin systems on a lattice 
such as the XY spin chain in a transverse magnetic field, it is possible to perform an exact analysis of the EE between two spatial blocks with the aid of the Fredholm determinant technique \cite{Its}. 
This technique based on the Riemann-Hilbert problem also plays a crucial role in the computation of the correlation functions for random matrices. On the other hand, it is known that the CS model is identical to Dyson's brownian motion model of the circular ensembles with $\beta =1,2,4$ \cite{Dyson}. 
Thus, it is promising to obtain the EE for spatial partitioning by applying the Fredholm determinant technique. It would also be interesting to investigate entanglement properties in integrable lattice models with inverse square interactions such as the Haldane-Shastry model \cite{Haldane-S,H-Shastry} and the long-range supersymmetric $t$-$J$ model \cite{Kuramoto-Yokoyama}. It remains an interesting issue whether 
our method developed in this article can be directly applied to these systems by using the {\it freezing trick} \cite{Polychronakos,Sutherland-Shastry}. 

\section{Acknowledgements}
The authors are grateful to
Y. Kato, S. Murakami, Y. Matsuo, R. Santachiara, and Y. Hatsugai for fruitful discussions.
This work was supported in part by Grant-in-Aids (Grant No. 15104006, No. 16076205, and No. 17105002) and NAREGI Nanoscience Project from the Ministry of Education, Culture, Sports, Science, and Technology.
HK was supported by the Japan Society for the Promotion of Science.

\appendix
\section{More detailed analysis}\label{app:detail}
In this appendix, we discuss the more detailed analysis of the EE (\ref{eq:EE}) and the universal subleading correction of the EE.  
Although both ${\cal S}_{N,L}$ and ${\cal S}_{N,L}^{\rm bound}$ go to infinity in the thermodynamic limit: $N \to \infty$ and $L$ is fixed,
the difference ${\cal S}_{N,L}^{\rm bound}-{\cal S}_{N,L}$ is finite.
The strategy to show this fact is to rewrite the sum over partitions as the integral over continuous variables.
This method is similar to the calculation of the dynamical correlation functions in the CS model by Lesage, Pasquier, and Serban \cite{Lesage}. 
Let us start with rewriting Eq.~(\ref{coeff_D}) in terms of parts of $\lambda$:
\begin{equation}
D_\lambda = \frac{1}{\binom{\beta N}{\beta L}} \prod_{j=1}^L \frac{\Gamma\big( \beta(L-j)+1 \big)\Gamma\big( \lambda_j+\beta(L-j+1) \big)}{\Gamma \big( \beta(L-j+1) \big)\Gamma \big( \lambda_j+\beta(L-j)+1 \big)}
=\frac{\beta^L L! \big( \beta(N-L) \big)!}{(\beta N)!} \prod_{j=1}^L \frac{\Gamma\big( \lambda_j+\beta(L-j+1) \big)}{\Gamma \big( \lambda_j+\beta(L-j)+1 \big)}.
\end{equation}
Introducing the new scaled variables $t_j=\lambda_j/N$ and using the Staring formula: $\Gamma(x+1)\overset{x\to \infty}{\sim} \sqrt{2\pi x} (x/e)^{x}$, we obtain the simple expression
for $D_\lambda$:
\begin{equation}
D_\lambda \sim \frac{L!}{N^L \beta^{(\beta-1)L}} f(t_1,\dots,t_L;\beta),
\end{equation}
where $f(t_1,\dots,t_L;\beta)=\prod_{j=1}^L t_j^{\beta-1}$.
In $N\to \infty$, we can replace the sum over \{$\lambda_j$\} with the integral over \{$t_j$\}:
\begin{equation}
\frac{1}{N^L} \sum_{0\le \lambda_L \le \cdots \le \lambda_1 \le \beta(N-L)} \to \int_D dt_1 \cdots dt_L,
\end{equation}
where $D$ is the region satisfying $0\le t_L \le \dots \le t_1 \le \beta$.
From these results, $\Tr \rho=\sum_\lambda D_\lambda$ is evaluated as
\begin{align}
\frac{L!}{\beta^{(\beta-1)L}}\int_D dt_1\cdots dt_L f(t_1,\dots,t_L;\beta)
&=\frac{1}{\beta^{(\beta-1)L}}\int_0^\beta dt_1 \cdots \int_0^\beta dt_L f(t_1,\dots,t_L;\beta) \nonumber \\
&=\frac{1}{\beta^{(\beta-1)L}} \left( \int_0^\beta dt\, t^{\beta-1} \right)^L =1.
\end{align}
This is consistent with the normalization condition of $\rho$.
Similarly, the EE ${\cal S}_{N,L}$ can be rewritten in terms of the integral over \{$t_j$\}:
\begin{align}
{\cal S}_{N,L}&=-\int_D dt_1\cdots dt_L \frac{L!}{\beta^{(\beta-1)L}}f \log \left( \frac{L!}{N^L \beta^{(\beta-1)L}}f \right) \nonumber \\
&=L\log N-\log L! +(\beta-1)L\log \beta-\frac{L!}{\beta^{(\beta-1)L}} \int_D dt_1\cdots dt_L f \log f.
\end{align}
We can exactly evaluate the integral of the last term as follows:
\begin{align}
\int_D dt_1\cdots dt_L f \log f&=
\frac{1}{L!}\int_0^\beta dt_1 \cdots \int_0^\beta dt_L (t_1 \cdots t_L)^{\beta-1} \sum_{j=1}^L \log t_j ^{\beta-1} \nonumber \\
&=\frac{L}{L!}\left(\int_0^\beta dt\, t^{\beta-1} \log t^{\beta-1}\right) \left( \int_0^\beta ds\, s^{\beta-1} \right)^{L-1} \nonumber \\
&=\frac{\beta^{(\beta-1)L}}{(L-1)!}(\beta \log \beta- \log \beta-1+\beta^{-1}).
\end{align}
Thus ${\cal S}_{N,L}=L\log N-\log L! +L(1-\beta^{-1})$.
On the other hand, since ${\cal S}_{N,L}^{\rm bound}=\log \binom{\beta(N-L)+L}{L} \sim L\log N-\log L!+L\log \beta$, we finally obtain
\begin{equation}
{\cal S}_{N,L}^{\rm bound}-{\cal S}_{N,L} \sim L(\log \beta-1+\beta^{-1}).
\label{difference}
\end{equation}
Therefore, the subleading term of the EE does not depend on the total number of particles $N$ but only on $L$ and the coupling of the CS model $\beta$. 
Note that the right hand side of Eq.~(\ref{difference}) vanishes only for $\beta=1$.
This result means the EE can saturate the upper-bound entropy only for the free fermion case.


\end{document}